# The Installation Costs of a Satellite and Space Shuttle Launch Complex as a Public Expenditure Project

**OZUYAR D.[1*], GUMUS OZUYAR S.[2], KARADENİZ O.[1] and VAROL Ö.[1]**

[1] Ankara University, Faculty of Science, Department of Astronomy and Space Sciences, 06100, Tandogan, Ankara, Turkey
[2] Hacettepe University, Department of Public Finance, 06800, Beytepe, Ankara, Turkey

*dozuyar@ankara.edu.tr

**Abstract**

From the 1940's to the present, space explorations, which is a highly important topic for the world and human beings, penetrate into many areas from the communication to the national security as well as from the discovery of exo-planets and new life forms to space mining. On the top of the countries which do researches on these fields are the developed countries and the developing countries are only used as launch areas, in an irrelevant manner of the research and development. However, developing countries can significantly reduce foreign dependency and security flaws as well as providing important reputation gain in international platforms by conducting space research and development activities as already done by developed countries. All the large-scale space probes conducted by developed countries oblige Turkey to develop space researches in terms of economy, security and scientific aspects. Due to these reasons, the approximate costs of a launch base, which will be installed to conduct space researches in Turkey, and of a satellite or a spacecraft, which will be able to launch from this base and serve a variety of purposes, are calculated in this study. In an effort to make the mentioned calculations, examples of various countries that have already established a launch base and already launched from these bases are analyzed and some projections are built for Turkey by calculating the estimated costs. Since these projections must be carried out by the Republic of Turkey since the private sector in Turkey will not be willing to invest in such activities, the possible public advantages can be gained through these activities are also mentioned and evaluated.

**Keywords :** Space sciences, Launching site, Public expenditure, Economic development, Turkey

The stars were used for different purposes by different civilizations in the human history. While ancient Egyptian civilization used the stars to determine the time of the Nile River flooding, Ancient Greek and Arab civilizations utilized them in the fields of mathematics and philosophy. By the help of these cultures' developments, Turk and Chinese civilizations conducted researches on space called as "the empyrean" and invented the first calendar concept with twelve animals (Unat, 2006:2). Although several improvements had been achieved in the earlier times mentioned above, the medieval time was the darkest period for the space practices because of the geo-centric universe model belief of church until Copernicus. Yet, the Europe's space interest raised again with the mathematical demonstration of the helio-centric universe model by Copernicus (Copernicus, 2002: 43), the assertion of the Earth's revolving around the Sun by Galilei and the discovery of elliptic planet orbits by Kepler (Cushing, 2003: 100-105). As a result of this enlightenment, the West gradually increased its attention on space explorations even the World Wars are experienced.

The space race, which had begun between Soviet Russia and the United States with the aim of providing military and political superiority, accelerated by Soviets' sending the first living being into space. As a result of the experiences gained from the different space missions with different species, the Soviet Union sent Yuri Gagarin into outer space as the first human in April 12, 1961. Gagarin returned into the Soviet Russian borders after his orbital flight around the Earth. Followingly, the US considerably increased the budget allocated for the space researches to keep up with the Russians, and National Aeronautics and Space Administration (NASA) was founded to carry out space explorations. Also, the US made a fundamental change in the education system and started investing in NASA with $89 million which was only 1% of the total expenditure budget in 1958 (Guardian Newspaper, 2010). After that, space exploration has become both an economic and technological race and a symbol of the development of a country.

Since sunk costs and also total fixed costs are economically quite high in this industry type, the private sector companies generally do not invest in it. Therefore, the space researches all over the world have mostly been carried out by state or with state assistance. In Turkey, this area does not draw considerable interest except for some research done in universities. Though there are several countries which do academic studies on the field of astronomy in the world, as of 2015, there are seventy countries with a government space agency (US Space Foundation, 2016). Of these countries, only thirteen have a facility with a launch capacity and seven of these sites are full-capacity launch areas (Japan Science and Technology Agency; Center for Research and Development Strategy, 2013). Except India, the countries having these sites are developed countries and the majority of the launch sites located in the developing countries are controlled and used by the developed countries. Therefore, the space expenditures of the developed countries are higher compared to those of the developing countries. According to the space economy report of the Organization for Economic and Co-operation and Development (OECD) in 2014, the space expenditures of the seven countries with a full launch capacity are approximately as follows; the US, $39.332 million; the People's Republic of China, $10.774 million; the Russian Federation, $8.691 million; the European Union, around $7.000 million; Japan, $3.421 million and Canada, $395 million (OECD, 2014).

India, which is the only country having a full launch capacity among the developing countries, is the only developing country heavily investing in the space explorations with the budget of $4.267 million (OECD, 2014). Even though Kazakhstan, another developing country, has a facility with a full launch capacity, it has rent all of the rights of this space base out to the Russian Federation (NASA, 2010) the outlays in this area are considered within the Russian Federation. Russia annually pays $115 million for lease and $50 million for facility maintenance and repair services to Kazakhstan for this base (IMF, 1995:1999). The outlays to the space sector in other developing countries are quite low; Brazil, $259 million; Indonesia, $142 million; Iran, $139 million; Turkey, $104 million; Israel, $89 million; South Africa, $76 million and Mexica, $8 million (Washington Post, 2014; OECD, 2014). These expenditures include the lease payments made to the countries with a launch base for construction and launching of communication and intelligence satellites.

Although the amount of the space expenditures varies depending on the development level of the country and the capacity utilization, the approximate costs of the launch areas having full capacity are quite close to each other. This is due to the similarities of the structures and facilities that should be in a launch base. In addition to the established settlements for the employees, depending on the nature of the satellite, a launch complex should contain at least an access tower, a launch pad, a launch mount, a mobile service tower, exhaust ducts, a shuttle assembly building, propulsion and fuel systems test buildings, clean rooms, a payload preparation room, a fuel holding area, liquid hydrogen and oxygen storage room as well as an emergency power building and a railway to transport the spacecraft to the launch pad (Arianespace, 2016).

Although these structures' costs are not completely itemized on the open sources, the total costs of some launch sites and space missions can be found from various sources. For example, the establishment expenditure of KouRou launch base located at 5 degrees North latitude is around 25 million French Franc in 1968, which is equivalent to 3 million and 750 thousand American Dollars with the exchange rate of that day (Russian Space Web, 2008). On the other hand, while the NASA's Juno mission launched in 2011 costs $1.1 billion (NASA, 2011), Russia will pay $3.4 billion for the Vostochny launch complex located at 51 degrees North latitude when the construction is completed (Russia Global Security, 2016). In this manner, Turkey has to allocate only 1,8% of its income to establish such a full launching facility with the same standard of Vostochny if the current exchange rate of dollar is assumed to be 2,96 TL. This budget share is smaller than the NASA's space budget between 1962 and 1969, in which NASA began to accelerate the space missions in order to get ahead of Russia.

Since the launch bases are natural monopoly production facilities, the profitability will gradually increase due to the cost reduction depending on its positive scale economy feature as production or launching is executed. Hence, even though the launch bases seem to have a high installation costs, the economic contributions of the launch bases are inevitably superior. For instance, India earned around $100 million from the satellite launching of 45 different countries (The Hindu newspaper, 2015). Also, NASA gained $1.93 billion revenue only in Florida in 2008 (NASA, 2008:1-2) and it annually brings $1 billion in only the technology transfer to the federal budget (NASA Socioeconomic Impacts, 2011).

Furthermore, when the historical progress of the countries conducted space researches are examined, it is observed that the public expenditures primarily raise the social capital as a result of the awareness, motivation and education and that physical capital and capacity are improved with this increment (US Information Office, 1959). With this irrevocable progress of a country, the capital increase in human and physical sense is actually one of the scales that determines the development level of that country. Due to the highness of the sunk costs such as installation and operating cost, the funding of the space researches has to be provided from the taxes since private sector will not be willing to invest in. Therefore, it is required to perform a comprehensive assessment concerning the public expenditure for the space researches since taxes of the society will be used for financing. In this context, a SWOT analysis was conducted with Ankara University astronomy students to have a better understanding on their opinions about a launch facility to be established in Turkey.

## Swot Analysis

**Strengths**
- Turkey is located between the latitudes of 36 and 42 degrees. It has a better latitudinal position than all other launch sites except KouRou and Cape Canaveral launch bases.
- Our close latitudinal position to the equator provides an advantage in terms of quickly escaping from the gravity, carrying heavier loads and using less fuel in space missions.
- Whereas the complicated orbital maneuvers are required in higher latitudes close to the poles, Turkey can obtain more effective results with less and easy orbital maneuvers due to its latitudinal location.
- Based on the geo-political position, Turkey can become the only potential launch base located in the junction of three continents.
- Turkey has a large number of young population that are graduated from limited number of universities with astronomy research departments and many of this population are unable to find any work in this field.
- Turkey can provide employees cheaper than developed countries.
- Turkey has the 18$^{th}$ powerful economy in the World.
- The presence of experienced Turkish astronomer such as İsmail Akbay who worked in a senior position at NASA between 1965 and 1973, and who was one of the chief engineers of Apollo projects.
- The experiences of the organizations such as Havelsan, Roketsan and Aselsan on aviation and satellite technologies.

**Weaknesses**
- The astronomical studies are limited to academic researches and application fields are inadequate.
- Turkish astronomers do not have enough engineering knowledge.
- Since the sunk costs which includes the installation costs are considerably high for this industrial area, it has to be made by state.
- Turkish astronomers are inexperienced in launching compared to their rivals.
- There are not enough educational and motivational notions encouraged the space activities.
- There is not any regulation about the conducting a domestic or foreign launch operations within country borders in Turkish legislations.
- With the 77 million population, Turkey's residential areas are quite dense and there is not quite possible to find unmanned large flat areas near the sea.
- Turkey is located on several fault lines which can produce intense earthquakes.

**Opportunities**
- The installation of a launch base eliminates the external dependence on some fields such as defense, communication and security since the land or software of other countries will not be used.
- When a satellite or a similar vehicle is launched in desired time, Turkey will take intelligence and defense advantages by getting information earlier than other resources.
- Since Turkey has a strong geo-political position, countries will highly prefer to use Turkey instead of other launching sites.
- The construction of a launch base and attraction of foreign investors will increase the economic gain.
- The development of new industries will increase employment opportunities by creating job opportunities in the country and hence this will reverse the brain drain.

- Acceleration and weight gain of the space researches will increase the quality and amount of the academic studies.
- Technological breakthroughs will affect the other economic fields and this will create an overall development wave.
- The presence of a launch base in Turkey will lead a power and prestige increment in foreign policy.

**Threats**

- Since the installation of a launch base in Turkey will cause reductions in both economic shares of other neighbor countries and the external dependency of Turkey, external pressures from foreign forces will highly increase.
- When the return travel of launched satellites to the Earth, they may cause environmental pollution by falling within the Turkish borders. On the other hand, if they fall out of Turkish borders, the technology may be passed into the hands of other countries.